\documentclass[useAMS,usenatbib,usegraphicx]{mn2e}

\usepackage{times,epsfig}

\def\gsim{\mathrel{\raise0.35ex\hbox{$\scriptstyle >$}\kern-0.6em
\lower0.40ex\hbox{{$\scriptstyle \sim$}}}}
\def\lsim{\mathrel{\raise0.35ex\hbox{$\scriptstyle <$}\kern-0.6em
\lower0.40ex\hbox{{$\scriptstyle \sim$}}}}
\def\m@th{\mathsurround=0pt }
\def\eqalign#1{\null\,\vcenter{\openup1\jot \m@th
 \ialign{\strut\hfil$\displaystyle{##}$&$\displaystyle{{}##}$\hfil
 \crcr#1\crcr}}\,}
\def\micron{$\mu\textrm{m}$ }
\def\micronend{$\mu\textrm{m}$}
\def\microjy{$\mu\textrm{Jy}$ }

\def\ergspers{ergs s$^{-1}$}

\title[Millimetre Survey of $z\sim 2$ ULIRGs]
      {A Millimetre Survey of Starburst Dominated Ultraluminous Infrared Galaxies at $z\sim 2$}
\author[Younger et al.]{
J.\,D.\ Younger,$^{\! 1}$\thanks{E-mail: jyounger@cfa.harvard.edu}
A.\,Omont,$^{\! 2}$
N.\,Fiolet, $^{\! 2}$
J.--S.\,Huang, $^{\! 1}$
G.\,G.\, Fazio, $^{\! 1}$ \and
K.\, Lai, $^{\! 1}$
M.\,Polletta, $^{\! 2,3}$
D.\, Rigopoulou, $^{\! 4}$
R.\, Zylka $^{\! 5}$
\vspace*{1mm}\\
$^1$ Harvard-Smithsonian Center for Astrophysics, 60 Garden Street,
     Cambridge, MA 02138, USA\\
$^2$ Institut d'Astrophysique de Paris, CNRS and Universit\'{e} Pierre et Marie Curie, 98bis boulevard Arago, 75014 Paris, France \\
$^3$ INAF-IASF Milano, via Bassini 15, 20133, Italy \\    
$^4$ Department of Astrophysics, Oxford University, Keble Road, Oxford, OX1 3RH, UK \\
$^5$ Institut de Radio Astronomie MillimŽtrique (IRAM), 300 rue de la Piscine, Domaine Universitaire, 38406 Saint Martin d'Hres, France
}

\date{\fbox{\sc Draft dated: \today\ }}
\pagerange{000--000}

\begin{document}

\maketitle

\begin{abstract}

We present millimetre observations of a sample of 12 high redshift ultraluminous infrared galaxies (ULIRGs) in the Extended Growth Strip (EGS).  These objects were initially selected on the basis of their observed mid--IR colours  ($0.0  < [3.6]-[4.5] < 0.4$ and $-0.7 < [3.6]-[8.0] < 0.5$) to lie at high redshift $1.5 \lsim z \lsim 3$, and subsequent 20--38\micron mid--IR spectroscopy confirms that they lie in a narrow redshift window centered on $z\approx 2$.  We detect 9/12 of the objects in our sample at high significance ($>3\sigma$) with a mean 1200\micron flux of $<F_{\rm 1200\mu m}> = 1.6\pm 0.1$ mJy.  Our millimetre photometry, combined with existing far--IR photometry from the Far--IR Deep Extragalactic Legacy (FIDEL) Survey and accurate spectroscopic redshifts, places constraints both sides of the thermal dust peak.  This allows us to estimate the dust properties, including the far--IR luminosity, dust temperature, and dust mass.  We find that our sample is similar to other high--$z$ and intermediate--$z$ ULIRGs, and local systems, but has a different dust selection function than submillimeter--selected galaxies.  Finally, we use existing 20cm radio continuum imaging to test the far--IR/radio correlation at high redshift.  We find that our sample is consistent with the local relation, implying little evolution.  Furthermore, this suggests that our sample selection method is efficient at identifying ultraluminous, starburst--dominated systems within a very narrow redshift range centered at $z\sim 2$.

\end{abstract}


\section{Introduction}

First discovered over 30 years ago, infrared (IR) luminous (LIRGs: $L_{\rm 8-1000\mu m} > 10^{11}$ $L_\odot$) and ultraluminous (ULIRGs: $L_{\rm 8-1000\mu m} > 10^{12}$ $L_\odot$) galaxies are among the most extreme objects in the universe, with energy outputs rivaling those of bright quasars \citep[see ][for a review]{sanders1996}.  Locally, theoretical modeling \citep{mihos1994} and both CO spectroscopic \citep[e.g.,][]{solomon1997} and optical/near--IR imaging \citep{veilleux2002} indicate that they are powered in large part by star formation (SF) induced by major mergers.  In the context of a merger--driven model of galaxy evolution, they represent the starburst prelude to the rapid, self--regulated growth of a nuclear supermassive black hole \citep[e.g.,][]{silk1998,murray2005,dimatteo2005,hopkins2007theory,younger2008.smbh}, a bright quasar, and eventually a ``red and dead" early--type massive galaxy \citep{sanders1988a,sanders1988b,hopkins2006,hopkins2007a,hopkins2007b}.  

Though locally they contribute a very small fraction of the infrared luminosity density, at high redshift $z\gsim 1$ LIRGs and ULIRGs take on increasing cosmological importance and may dominate cosmic SF at $z\gsim 2$ \citep{blain1999,blain2002,lefloch2005}.  There has also been recent observational evidence for a significant populations of hyperluminous sources at still greater redshift $z\gsim 4$ \citep{wang2007,younger2007,younger2008,dannerbauer2008,younger2008highres}.  Several techniques have been developed for selecting high redshift ULIRGs, including: direct far--IR selection via (sub)millimeter surveys \citep[e.g.,][]{hughes1998,barger1998,greve2004,pope2006,scott2008}, dust--obscured galaxies (DOGs) selected on the basis of their 24\micronend--R colour \citep[e.g.,][]{houck2005,yan2005}, and those selected based on their mid--IR colours \citep[][Huang et al., in preparation]{farrah2008}.  Understanding the nature of these different populations, and the engine driving their extreme luminosities, is crucial to a thorough understanding of SF and galaxy evolution at high redshift.   

\begin{table*}
\begin{tabular}{cccccccc}
\hline
\hline
Object & Name & r' & $F_{\rm 24\mu m}$ & $F_{\rm 70\mu m}$ & $F_{\rm 160\mu m}$ & $F_{\rm 1200 \mu m}$ & $F_{\rm 20 cm}$ \\
& & [mag] & [$\mu$Jy] & [mJy] & [mJy] & [mJy] & [$\mu$Jy] \\
\hline
EGS142301.5+523223 & egs1 & $23.18\pm 0.02$ & $554\pm 35$ & $<3.0$ & $12.7\pm 7.0$ & $1.19\pm 0.39$ & $59\pm 10$ \\
EGS142148.5+521435 & egs4 & $26.58\pm 0.20$ & $557\pm 22$ & $2.4\pm 1.0$ & $<21.0$ & $1.65\pm 0.39$ & $68\pm 10$ \\
EGS141928.1+524342 & egs10 & $23.32\pm 0.03$ & $623\pm 35$ & $5.0\pm 1.0$ & $45.5\pm 7.0$ & $1.31\pm 0.36$ & $85\pm 10$ \\
EGS141917.4+524922 & egs11 & $26.37\pm 0.21$ & $591\pm 20$ & $8.7\pm 1.0$ &  $<21.0$ & $0.84\pm 0.40$ & $67\pm 10$ \\
EGS141920.4+525038 & egs12 & $23.43\pm 0.02$ & $743\pm 23$ & $<3.0$ & $<21.0$ & $1.57\pm 0.41$ & $36\pm 10$ \\
EGS141900.3+524948 & egs14 & $23.83\pm 0.06$ & $1053\pm 41$ & $4.5\pm 1.0$ & $76.7\pm 7.0$ & $3.70\pm 0.49$ & $263\pm 10$ \\
EGS141829.5+523630 & egs21 & $24.16\pm 0.05$ & $605\pm 23$ & $2.5\pm 1.0$ & $35.2\pm 7.0$ & $1.11\pm 0.30$ & $70\pm 10$ \\
EGS141822.5+523938 & egs23 & $25.06\pm 0.07$ & $665\pm 18$ & $2.9\pm 1.0$ & $62.4\pm 7.0$ & $1.80\pm 0.38$ & $119\pm 10$ \\
EGS141834.6+524506 & egs24 & $24.25\pm 0.05$ & $663\pm 29$ & $4.4\pm 1.0$ & $9.7\pm 7.0$ & $0.36\pm 0.38$ & $47\pm 10$ \\
EGS141746.2+523322 & egs26 & $25.84\pm 0.24$ & $492\pm 16$ & $1.5\pm 1.0$ & $21.6\pm 7.0$ & $1.13\pm 0.35$ & $97\pm 10$ \\
EGS141836.8+524604 & egs24a & $24.56\pm 0.17$ & $997\pm 30$ & $2.5\pm 1.0$ & $15.1\pm 7.0$ & $2.85\pm 0.47$ & $112\pm 10$ \\
EGS142219.8+531950 & egsb2 & $23.74\pm 0.03$ & $616\pm 30$ & $3.6\pm 1.0$ & $21.7\pm 7.0$ & $-0.57\pm 0.54$ & $134\pm 10$ \\
\hline
\end{tabular}
\caption{The observed r' band fluxes and far--IR SEDs of the ULIRGs in our sample.  Upper limits are at the $3\sigma$ level.  The 70\micronend,160\micronend, and 20 cm photometry is from Huang et al. (in preparation).  The 1200\micron MAMBO observations and data reduction are described in \S~\ref{sec:obs}.  }
\label{tab:sources}
\end{table*}

At the same time, a remarkably tight correlation between the radio and far--IR luminosity of galaxies, spanning several decades in FIR luminosity, has been known for more than 20 years \citep{helou1985,condon1991,condon1992,yun2001}.  Though it is thought that this relationship arises from synchrotron loses associated with massive star forming regions \citep{hummel1986,hummel1988,condon1992}, the detailed physics are relatively poorly understood \citep[see also the discussion in][]{thompson2006}.  Therefore, the persistence of this relation at intermediate to high redshift -- where the conditions of galaxy formation and star forming environment were likely quite different than they are locally -- is a subject of great interest in current research \citep{garrett2002,gruppioni2003,appleton2004,kovacs2006,boyle2007,vlahakis2007,ibar2008,sajina2008}.

In this work, we present far--IR observations of a sample of 12 mid--IR selected ULIRGs at $z\sim 2$, first identified by Huang et al. (in preparation).  Our far--IR photometry, which spans both sides of the thermal peak, allows us to more robustly estimate the luminosity and dust properties -- e.g., dust temperature and mass -- of these objects.  The relative uniformity of this sample, combined with mid--IR spectroscopic redshifts that indicate a remarkably tight redshift range, makes it ideal for such a study.  It furthermore provides a promising platform for probing the far--IR/radio correlation at $z\sim 2$.  

\begin{figure}
\epsfig{figure=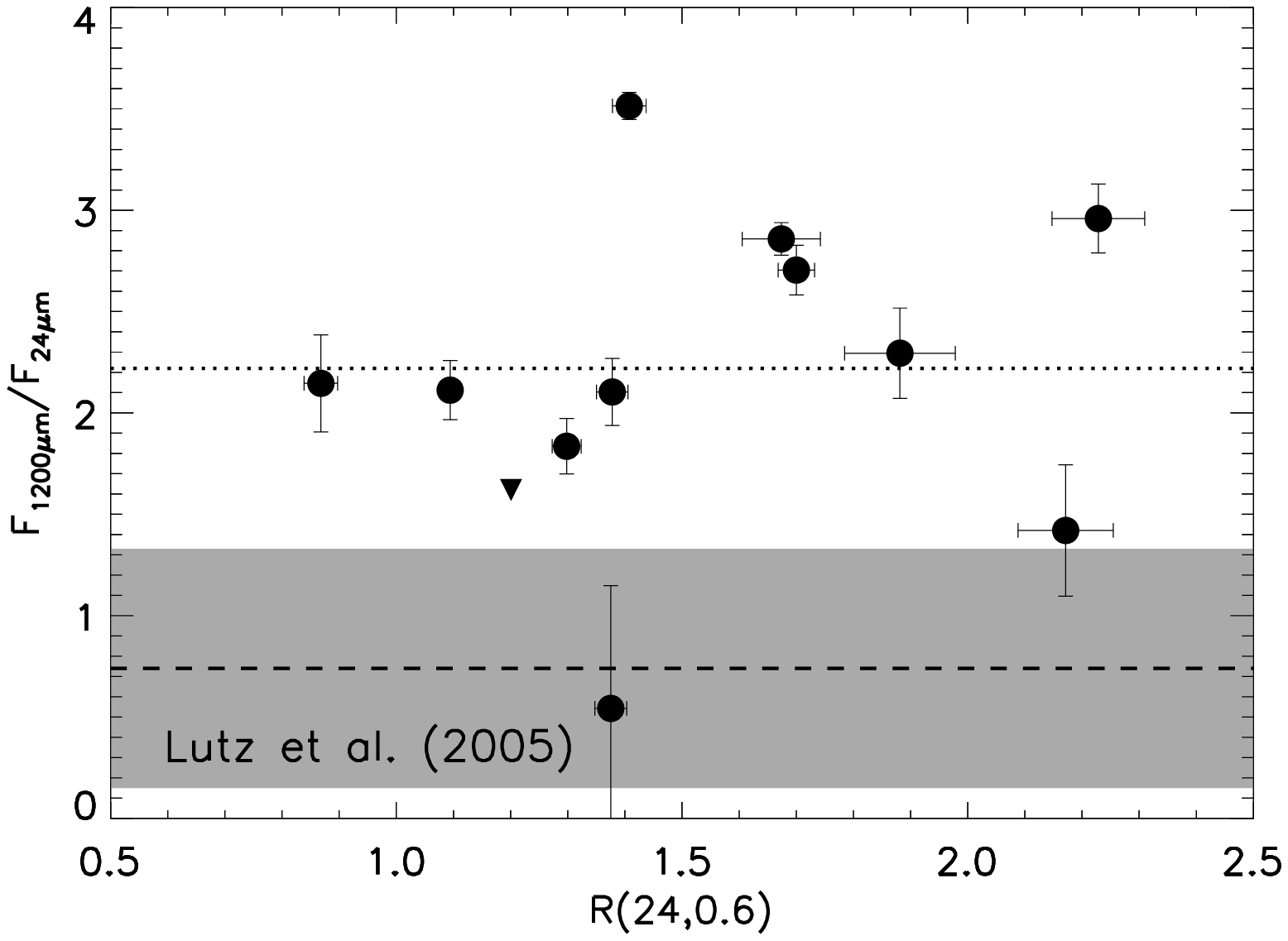,width=79mm}
\caption{The correlation between the millimetre--to--24\micron flux density ratio ($\tilde{R}(1200,24) = S_{\rm 1200\mu m}/S_{\rm 24\mu}$) and the obscuration parameter of \citet[][]{lutz2005}: $R(24,0.6) = {\rm log_{10}}(\nu F_\nu({\rm 24 \mu m})/\nu F_\nu({\rm 0.6 \mu m}))$.  Upper limits are at the $3\sigma$ level are indicated by inverted, filled triangles.  We find the all of our sources fall above the mean measured by \citet[shaded region and dashed line]{lutz2005}, with a mean of $<\tilde{R}(1200,24)> = 2.22\pm 0.08$ (dotted line) with an intrinsic dispersion of $\sigma_{\tilde{R}(1200,24)} = 0.83$.  Therefore, our sources are more far--IR luminous at fixed 24\micron independent of obscuration, which is consistent with a larger contribution from a starburst starburst over a dusty AGN.}
\label{fig:r24r}
\end{figure}

This paper is organized as follows: in \S~\ref{sec:sample} we outline our sample selection, in \S~\ref{sec:obs} we detail our observations, in \S~\ref{sec:results} we present the results of our observations, in \S~\ref{sec:fit} we describe our far--IR spectral energy distribution fitting method, in \S~\ref{sec:dust.properties} we present estimates of the dust properties for objects in our sample, in \S~\ref{sec:fir.radio} we examine the far--IR/radio correlation at high redshift, and in \S~\ref{sec:conclusion} we conclude.  Throughout this work, we assume a flat concordance cosmology with $\Omega_m = 0.3$, $\Omega_\Lambda = 0.7$ and $h=0.7$ \citep{spergel2003,spergel2007,tegmark2004,tegmark2006}.  

\section{Sample Selection}
\label{sec:sample}

It has been shown \citep{sawicki2002,huang2004,papovich2008} that an efficient method for selecting high redshift galaxies is to track the 1.6\micron stellar emission bump via mid--IR colors.  Using the Infrared Array Camera \citep[IRAC:][]{fazio2004} on board the {\em Spitzer} Space Telescope \citep{werner2004}, applying the color selection criteria
\begin{eqnarray}
& 0.0  & < [3.6]-[4.5] < 0.4 \,\,\,\, {\rm and} \\
& -0.7& < [3.6]-[8.0] < 0.5,
\end{eqnarray}
where [3.6] denotes the AB magnitude in the 3.6\micron band (and likewise for the 4.5 and 8.0\micron bands), will identify massive galaxies in the redshift range $1.5\lsim z \lsim 3.0$ -- the redshift range where the peak of the 1.6\micron stellar bump is redshifted into the 4.5 and 5.8\micron bands.  Furthermore, since this is effectively a rest frame near--IR selection criterion, it is relatively insensitive to dust reddening and is roughly stellar--mass limited \citep{huang2004,huang2005,conselice2007a}.  

Huang et al. (in preparation) used this technique, combined with a 24\micron flux limit of $F_{\rm 24\mu m}\geq 0.5$ mJy using the Multiband Imaging Photometer for Spitzer \citep[MIPS:][]{rieke2004}, to select high redshift, massive, IR luminous systems for mid--IR spectroscopic followup using the Infrared Spectrograph \citep[IRS:][]{houck2004} -- both also on board {\em Spitzer}.  This yielded a sample of 12 objects in the Extended Groth Strip (EGS), all with $L_{IR} \gsim 10^{12} L_\odot$ in a remarkably tight redshift range centered around $z \sim 1.9$.  This was also a considerably more uniform sample of high redshift IR luminous galaxies than other {\em Spitzer} selected samples; see Huang et al. (in preparartion) for a more detailed comparison to the various other selection techniques.  Our sample was drawn from the Huang et al. (in preparation) EGS sample with mid--IR derived spectroscopy redshifts.

\section{Observations}
\label{sec:obs}

The optical and near-- to mid--IR observations used for this work are described in detail in \citet{davis2007}, Huang et al. (2008, in preparation), and references therein.  Briefly, we make extensive use of deep $r'$ imaging with a limiting magnitude of 26.5 AB, IRAC imaging at (3.6, 4.5, 5.8, 8.0)\micron to $5\sigma$ flux limits of (1,1, 1.2, 6.3, 6.9)\microjy \citep[see also][]{barmby2004,barmby2008}, and MIPS imaging at 24\micron to a $5\sigma$ depth of 77\microjy \citep{sanders2007}.  We also make use of extremely deep far--IR MIPS imaging at 70 and 160\micron that combined existing observations (PID 00008, PI Fazio) and the Far--Infrared Deep Extragalactic Survey (FIDELS; PID 30948, PI Dickinson)\footnote{For a description of the FIDELS observing strategy and data reduction, see http://data.spitzer.caltech.edu/popular/fidel/2007\_sep17/fidel\_dr2.html.}, and deep 1.4 GHz radio imaging to a depth of 50\microjy \citep{ivison2007c}.  Photometry for our sources was first presented by Huang et al. (2008, in preparation).

\begin{figure*}
\epsfig{figure=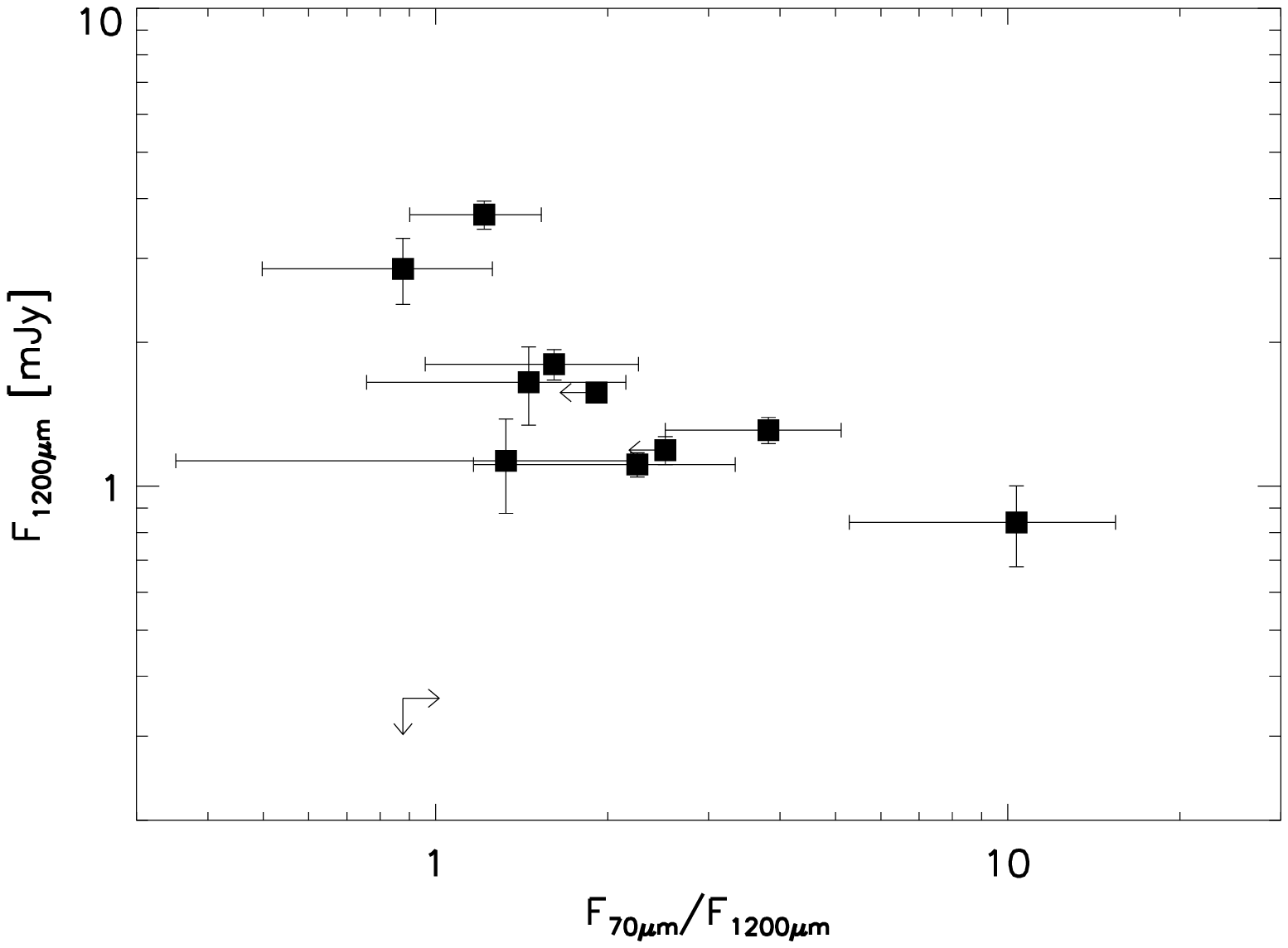,width=75mm}
\epsfig{figure= 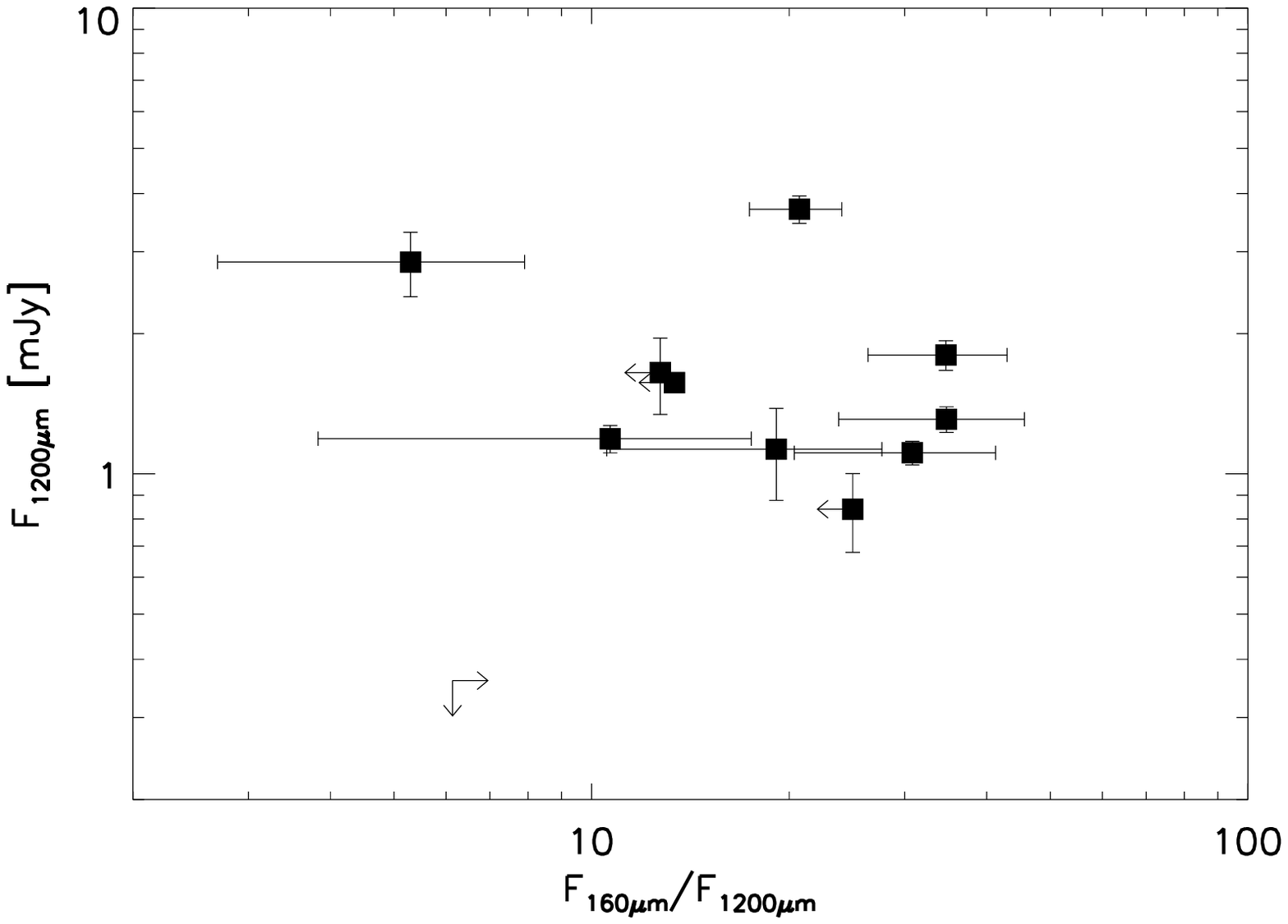,width=75mm}
\caption{The millimetre detectability as a function of the 70\micronend-- ($F_{\rm 70\mu m}/F_{\rm 1200\mu m}$: left) and 160\micronend--to--millimetre ($F_{\rm 160\mu m}/F_{\rm 1200\mu m}$: right) flux density ratio.  Upper and lower limits are indicated by arrows.  We find a strong negative correlation at the $\sim 2\sigma$ level (see Table~\ref{tab:pah.rank}) between the millimetre flux and $F_{\rm 70\mu m}/F_{\rm 1200\mu m}$, reflecting the fact that warmer sources are less likely to have a significant cold dust component.  We do not, however, find a significant correlation between the millimetre flux and $F_{\rm 160\mu m}/F_{\rm 1200\mu m}$.}
\label{fig:rdust}
\end{figure*}

The millimetre observations presented here were performed at the Institut de Radioastronomie Millimetrique (IRAM) 30 metre telescope at Pico Veleta during the Winter/Spring of 2008, using the 117 element Max--Plank Bolometer Array 2 \citep[MAMBO2:][]{kreysa1998}.   All our observations were performed in good weather conditions, with $\tau_{\rm 230 GHz} \lsim 0.3$ and 1.9 Hz (corresponding to a 0.52s wobbler ``period") chopped sky noise $\lsim 100$ mJy/beam (HPBW=11 arcsec).  Primary pointing, focus, and flux calibration was performed on either Mars ($\sim 130$ Jy) or another bright calibrator source -- including K3--50A ($\sim 7$ Jy) and CW--LEO ($\sim 1.3$ Jy).  Secondary pointing was performed on J1419+544 ($\sim 1$ Jy), a bright quasar only 1 degree away.  On--off observations were obtained in 20 minute scans with a wobbler throw of 35 arcseconds, which were repeated until either a significant detection (S/N $\gsim 3\sigma$) or a noise level of $\sim 0.4$ mJy.  This typically necessitated $\sim 2$ hours of integration time per source (full range 1--3 hours).  The data were reduced using the {\sc MOPSIC} Package developed by R. Zylka, using the standard reduction pipeline.  

\section{Results}
\label{sec:results}

In all, 9 of the 12 targets had significant detections ($\gsim 3\sigma$) at 1200\micronend -- egs11 was detected at $\sim 2\sigma$, egs24 was detected at $\sim 1\sigma$, and egsb2 was consistent with the sky noise.   Excluding egsb2, the mean observed flux density $<F_{\rm 1200\mu m}> = 1.6\pm 0.1$ mJy and mid-- to far--IR colour $<\tilde{R}(1200,24)> = <S_{\rm 1200\mu m}/S_{\rm 24\mu m}> = 2.22\pm 0.08$ with an intrinsic dispersion of $\sigma_{\tilde{R}(1200,24)} = 0.83$.  The average 1200\micron flux density and detection rate is similar to the IRAC selected starburst sample \citet{lonsdale2008}.  This average flux density ratio noticeably lower than is typical of bright SMGs \citep[$<\tilde{R}(1200,24)>\approx 12$:][]{egami2004,ivison2004}.  However, while it is consistent with the detected sources in their sample, this is a significantly higher average value than has been observed in other {\em Spitzer} selected samples of high redshift ULIRGs: \citep{lutz2005} followed up a sample of 40 obscured 24\micron sources \citep[see also][]{houck2005,yan2005,weedman2006a,weedman2006c} with MAMBO at 1.2mm and found an overall $<\tilde{R}(1200,24)> = 0.74\pm 0.09$ with $\sigma_{\tilde{R}(1200,24)} = 0.59$, which is consistent with the expectation from an IR bright active galactic nucleus \citep[AGN:][]{elvis1994}.  \citet{sajina2008} found similar millimetre fluxes for a comparable sample \citep{yan2007}.  This is not unexpected, as obscured 24\micron sources are thought to contain a large fraction of AGN \citep{weedman2006c}.

In Figure~\ref{fig:r24r}, we show the obscuration parameter from \citet{lutz2005}, $R(24,0.6) = {\rm log_{10}}(\nu F_\nu({\rm 24 \mu m})/\nu F_\nu({\rm 0.6 \mu m}))$, where $\nu F_\nu({\rm 0.6 \mu m})$ is taken from the $r'$ imaging \citep{davis2007}.  These authors found that when they restricted their sample to the most obscured sources with $R(24,0.6) \geq 1.5$, the average $\tilde{R}(1200,24)$ increased to $0.74\pm 0.09$ with $\sigma_{\tilde{R}(1200,24)} = 0.59$.  Though they could not examine it directly due to the low detection rate of sources in their sample, this suggests a correlation between $R(24,0.6)$ and $\tilde{R}(1200,24)$.  For our sources with ${\rm S/N} \geq 2$, a Spearman rank correlation coefficient analysis finds a mild correlation with $\rho = 0.27$ at the $\sim 1\sigma$ level.  The low significance of this correlation may be due to small sample size, but it is broadly consistent with the findings of \citet{lutz2005} and suggests that the most heavily obscured sources efficiently reprocess the attenuated stellar radiation into cold ($\lsim 60$ K) thermal dust emission.

Given the narrow redshift range of our sources, we can also use the 70\micron and 160\micronend--to--millimetre flux density ratios to investigate the correlation between strong millimetre emission and the dust properties of objects in our sample.  At $z\sim 2$, the 70 and 160\micron MIPS channels probe the rest--frame 24 and 50\micron emission.  These wavelengths probably probe AGN--heated dust \citep[e.g.,][]{elvis1994,urry1995} and cold dust from a starburst \citep{sanders1996} respectively.  In Figure~\ref{fig:rdust} we show the millimetre flux as a function of both $F_{\rm 70\mu m}/F_{\rm 1200\mu m}$ and $F_{\rm 160\mu m}/F_{\rm 1200\mu m}$ respectively, and in Table~\ref{tab:pah.rank} we present the results of a Pearson rank correlation coefficient analysis.  We find a strong negative correlation between the millimetre flux and $F_{\rm 70\mu m}/F_{\rm 1200\mu m}$; sources with more hot relative to cold dust -- and therefore likely a larger AGN contribution -- are less likely to be detected at 1200\micronend.  We do not however, find a strong correlation between $F_{\rm 160\mu m}/F_{\rm 1200\mu m}$ and $F_{\rm 1200\mu m}$.  We believe that this is mostly likely a result of the relatively poor sensitivity of the 160\micron FIDELS survey: if an object has enough cold dust to be detected at 160\micron at all, it is also likely to be a strong millimetre source.

\begin{table}
\begin{tabular}{cccccccc}
\hline
\hline
Parameter & $\rho$ & Significance \\
& & [$\sigma$] \\
\hline
$F_{\rm 70\mu m}/F_{\rm 1200\mu m}$ & -0.76 & 2.0 \\
$F_{\rm 160\mu m}/F_{\rm 1200\mu m}$ & -0.10 & 0.3 \\
$L_{\rm 7.7\mu m}$ & -0.04 & 0.1 \\
$L_{\rm 11.3 \mu m}$ & 0.26 & 0.8 \\
${\rm EW}_{\rm 7.7\mu m}$ & -0.62 & 1.6 \\
${\rm EW}_{\rm 11.3\mu m}$ & 0.31 & 0.7 \\
\hline
\hline
\end{tabular}
\caption{The results of a Spearman rank correlation coefficient ($\rho$) analysis of the trends of different PAH parameters (see Figure~\ref{fig:pah} with the observed millimeter flux density $F_{\rm 1200 \mu m}$.}
\label{tab:pah.rank}
\end{table}

\begin{figure*}
\epsfig{figure=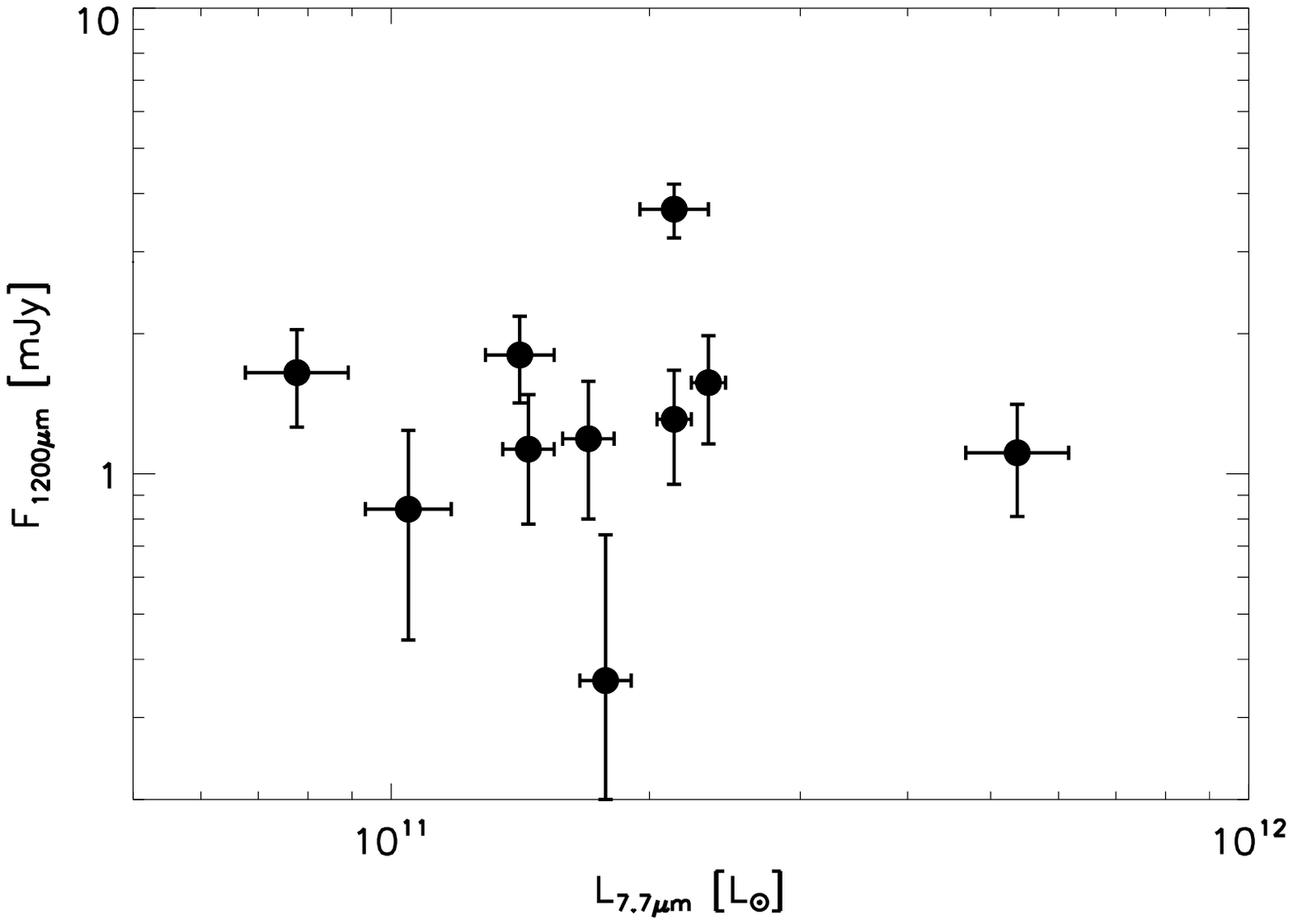,width=79mm}
\epsfig{figure=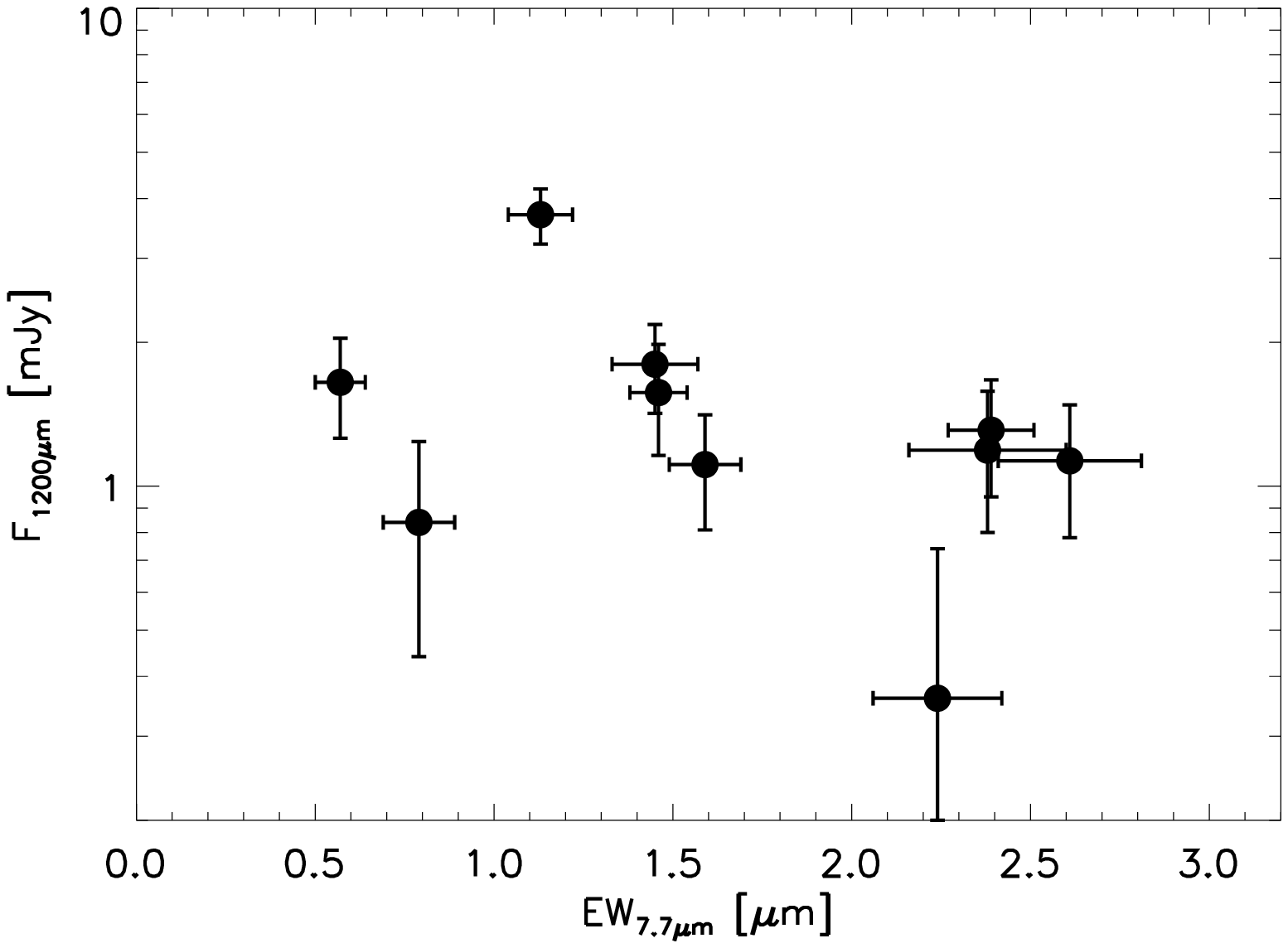,width=79mm}
\epsfig{figure=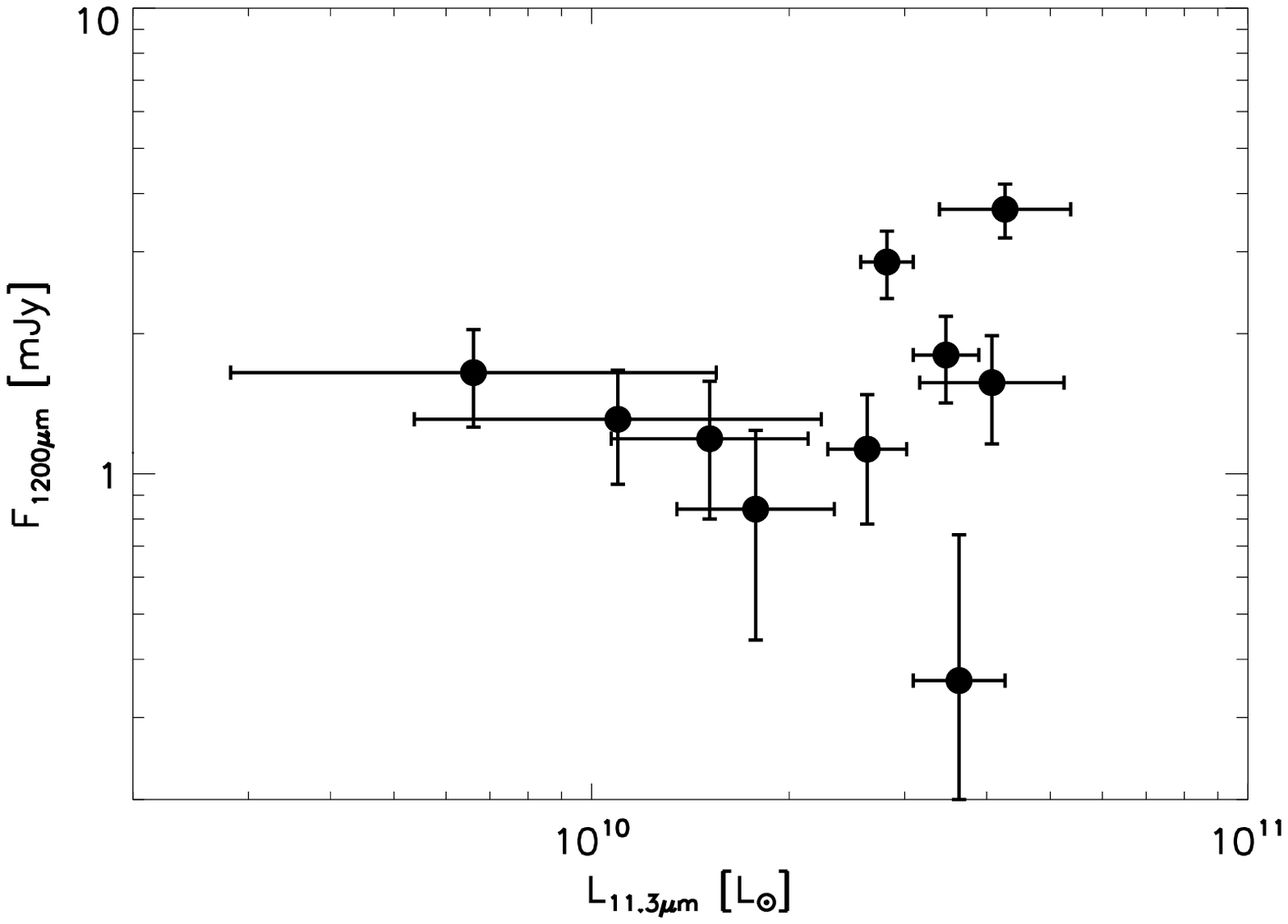,width=79mm}
\epsfig{figure=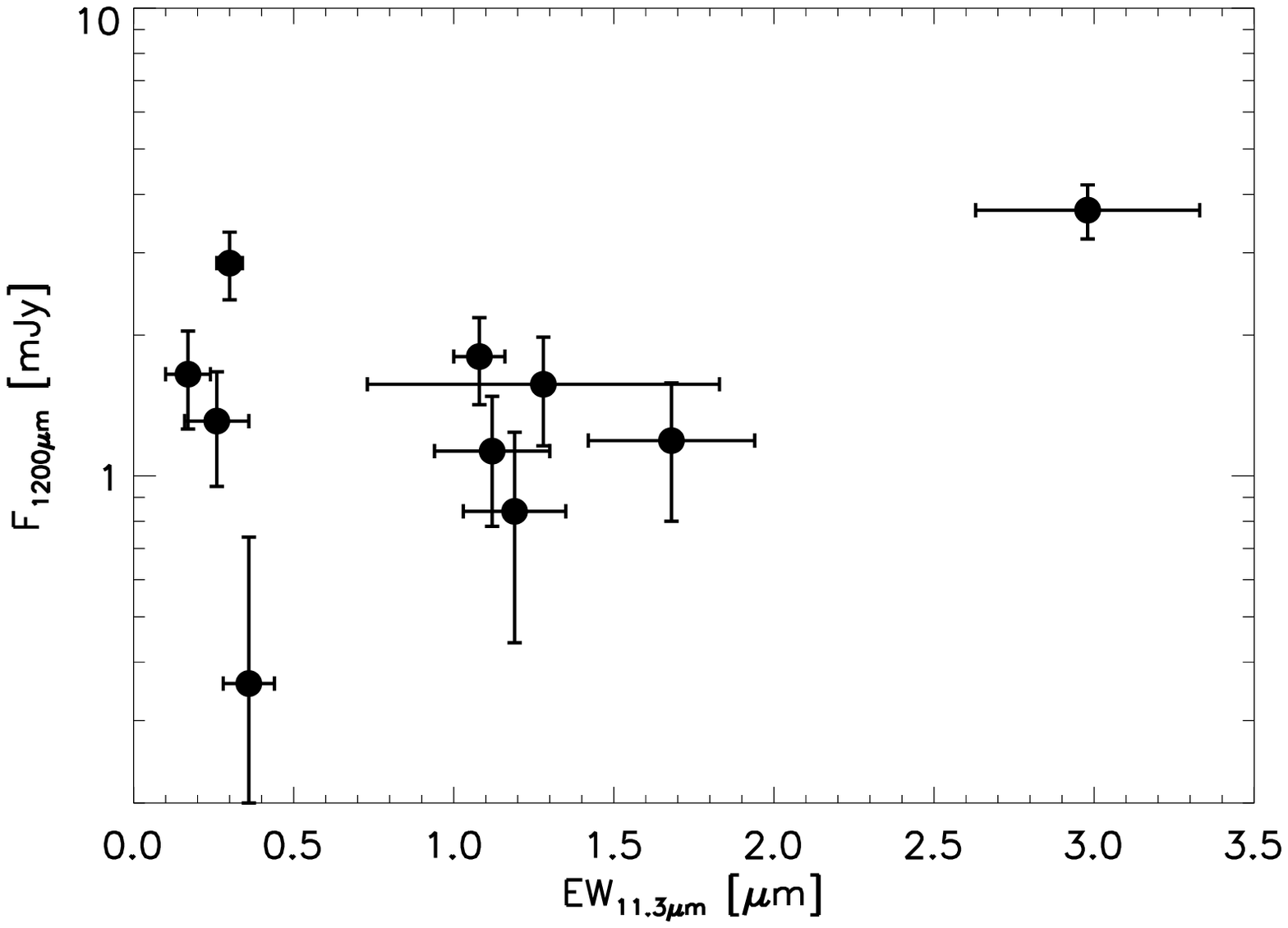,width=79mm}
\caption{The correlation between the observed millimeter flux density $F_{\rm 1200\mu m}$ and PAH luminosities (left: $L_{\rm 7.7\mu m}$ and $L_{\rm 11.3\mu m}$) and equivalent widths (right: ${\rm EW}_{\rm 7.7\mu m}$ and ${\rm EW}_{\rm 11.3\mu m}$) at 7.7 (top) and 11.3\micron (bottom).  The results of a Pearson rank correlation coefficient analysis are shown in Table~\ref{tab:pah.rank}.}
\label{fig:pah}
\end{figure*}

It been observed that the PAH luminosity is correlated with the far--IR \citep{rigopoulou1999,wu2005,brandl2006,calzetti2007,desai2007,shi2007}, and therefore we would expect strong PAH sources to show strong millimetre emission.  To investigate this for our sample, which contains all strong PAH emitters, in Figure~\ref{fig:pah} we show the correlations between the millimetre flux ($F_{\rm 1200\mu m}$) and PAH luminosities ($L_{\rm 7.7\mu m}$ and $L_{\rm 11.3\mu m}$) and equivalent widths (${\rm EW}_{\rm 7.7\mu m}$ and ${\rm EW}_{\rm 11.3\mu m}$) at 7.7 and 11.3\micronend.  The results of a Spearman rank correlation coefficient analysis are presented in Table~\ref{tab:pah.rank}.  All of these results are low significance, likely due to the effects of a relatively small and homogenous sample.  They are broadly consistent -- considering the low significance of the $L_{\rm 7.7\mu m}$ inverse correlation -- with observations of local systems that found a correlation between the PAH luminosity and ongoing star formation \citep[e.g.,][]{wu2005,brandl2006,calzetti2007}.  In particular, we find a positive correlation between 11.3\micron PAH luminosity and equivalent width and the millimetre flux at a $0.8$ and $0.7\sigma$ level of significance respectively.  Since the 11.3\micron feature is more sensitive to obscuration \citep{rigopoulou1999}, sources with strong observed features at this wavelength are likely to be more intense starbursts, and thus more likely to be detected in the far--IR.  However, we find a significant negative correlation between the 7.7\micron equivalent width and $F_{\rm 1200\mu m}$ at $1.6\sigma$ significance.  This differs from the results of \citet{sajina2008} who find a positive correlation ($\rho=0.56$) between $F_{\rm 1200\mu m}$ and ${\rm EW}_{\rm 7.7\mu m}$ at $\sim 3\sigma$ confidence for a sample of high redshift obscured {\em Spitzer} selected ULIRGs \citep{yan2007}.  While the correlation seen for our sources is, again, low--significance, it suggests that the relationship between strong PAH emission and far--IR luminosity is somewhat different in their sample -- a reasonable result considering the far different mid-- to far--IR properties of similar objects \citep{lutz2005} are significantly different from those presented here. 

\section{Spectral Energy Distribution Fitting}
\label{sec:fit}

\begin{figure*}
\epsfig{figure=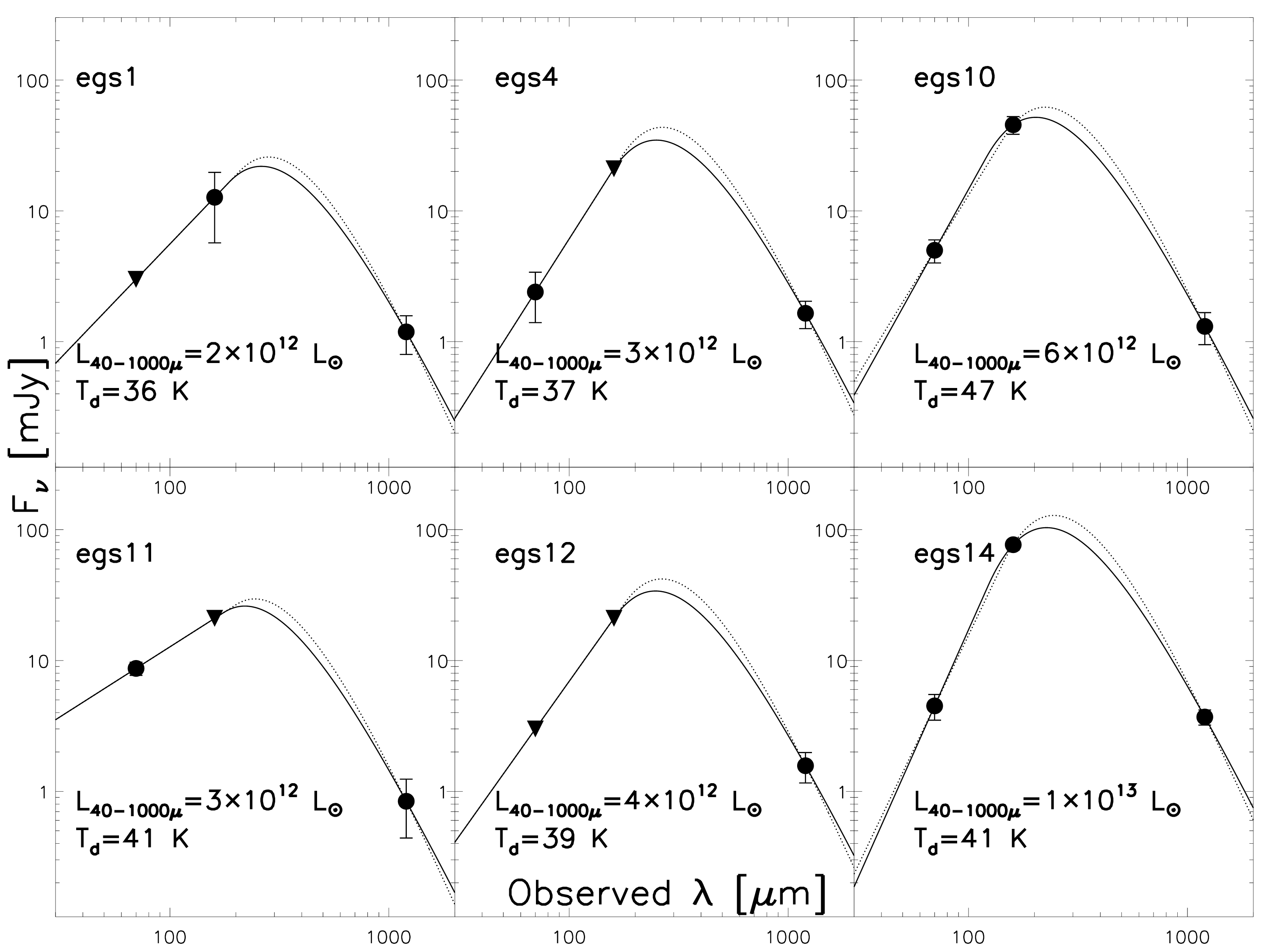,width=150mm}
\caption{The observed far--IR SEDs (see Table~\ref{tab:sources}) of the $z\sim 2$ ULIRGs in our sample, along with the fitted model.  The model parameters and derived quantities are listed in each panel, as well as tabulated in Table~\ref{tab:fits}, and the fitting procedure summarized in \S~\ref{sec:fit}.  The solid lines have a fixed dust emissivity $\beta=1.5$, while the dotted lines are the same for $\beta=2.0$.}
\label{fig:fits1}
\end{figure*} 

\begin{figure*}
\epsfig{figure=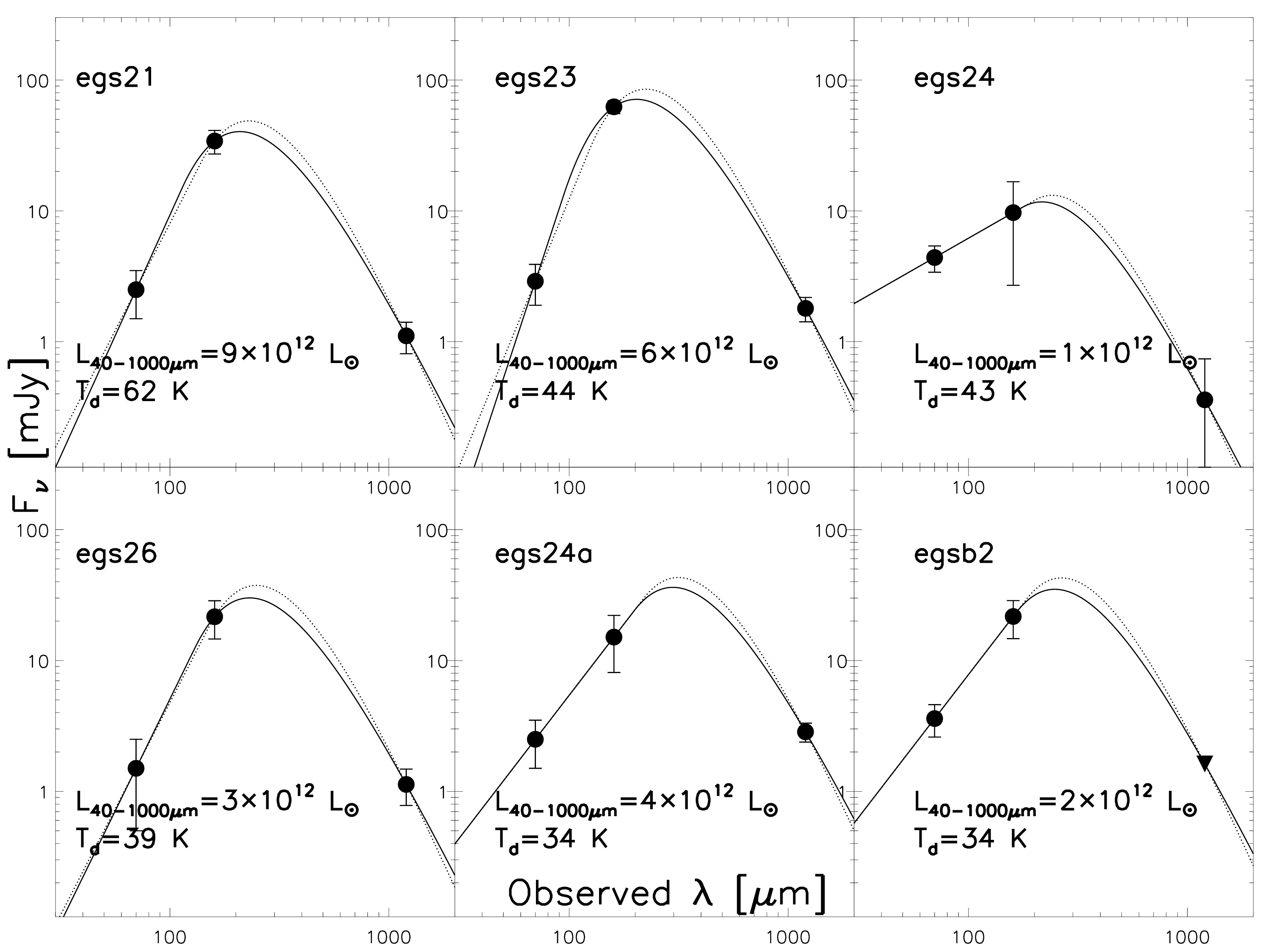,width=150mm}
\caption{Same as Figure~\ref{fig:fits1}.}
\label{fig:fits2}
\end{figure*} 

Our millimetre observations constrain the shape of the far--IR SED of our sources with photometry on both sides of the thermal dust peak.  Given a choice of a simple far--IR SED parameterization, this allows us to characterize the dust properties in a sample of very luminous, high redshift systems from the far--IR directly, without invoking additional assumptions -- for example assuming the local far--IR/radio correlation \citep[e.g.,][]{chapman2005}.  For this work, we use a single temperature greybody fitting form first introduced by \citet{hildebrand1983}, in which the thermal dust spectrum is approximated as $F_\nu = Q_\nu B_\nu(T_d)$, where $B_\nu$ is the Planck function, $Q_\nu = Q_0 (\nu/\nu_0)^\beta$ is the dust emissivity with $1 \lsim \beta \lsim 2$, and $T_d$ is the effective -- or emission weighted -- dust temperature.  This yields a dust spectrum that for $h \nu/k T_d \gsim 1$ takes the form:
\begin{equation}
\label{eq:greybody}
F_\nu \propto \frac{\nu^{3+\beta}}{exp(h\nu/k T_d)-1}
\end{equation}
This is then typically paired with a power--law component at short wavelengths $F_\nu \sim \nu^{-\alpha}$, to account for a subdominant warmer dust component arising either from the starburst or an AGN, which is matched to the long wavelength SED at $\partial F_\nu/\partial\nu = -\alpha$ to ensure a continuous first derivative.  \citet{blain2003} find that this form provides a good match to the SEDs of observed systems from $\lambda \sim 30-1000$\micron over a large range in total IR luminosities \citep[see also][for additional discussion of far--IR fitting methods]{blain1999a,carilli1999,dunne2000,yun2002,sajina2006,yang2007a}

This model was fit to the 70, 160, and 1200\micron data using a Levenberg-Marquardt least-squares minimization routine, the results of which are summarized in Figures~\ref{fig:fits1} and \ref{fig:fits2} and Table~\ref{tab:fits}.   Since $T_d$ is degenerate with $\beta$ \citep{blain2003,sajina2006} for sparsely sampled far--IR SEDs, we fix $\beta = 1.5$ which is consistent with the results of SED fitting of local and high redshift systems \citep{dunne2000,yun2002,yang2007b}; using a higher value of $\beta$ will lead to systematically lower dust temperatures \citep{blain2003,sajina2006}.  Furthermore, the accurate redshifts supplied by mid--IR spectroscopy were essential in removing the additional degeneracy between $T_d$ and redshift \citep{carilli1999,yun2002,blain2003}.  From these fitted models we can estimate $T_d$ and the far--IR luminosity.  We can also estimate the total dust mass in these systems according to:
\begin{equation}
\label{eq:dustmass}
M_d = \frac{S_\nu D_L^2}{\kappa(\lambda_{\rm rest}) B_\nu(\lambda_{\rm rest},T_d)}
\end{equation}
where $M_d$ is the total dust mass, $S_\nu$ is the observed flux density, $D_L$ is the luminosity distance, $\kappa(\lambda_{\rm rest})$ is the rest frame dust mass absorption coefficient at the observed wavelength \citep[for which we use the Milky Way dust model of][]{weingartner2001}, and $B_\nu(\lambda_{\rm rest},T_d)$ is the Planck function at the rest wavelength.  At long wavelengths -- e.g., 1200\micron observed, $\sim 400$\micron in the rest frame -- observations are directly probing populations of dust that are preferentially colder than $T_d$, which is effectively an emission weighted average over dust populations at a large range of physical temperatures.  Therefore, we use the fitted models to estimate the peak flux and wavelength, which is dominated by dust at $T_d$ and simultaneously minimizes the effects of uncertainty in $\beta$ ($\kappa_\nu \propto \nu^{-\beta}$), for input into equation \ref{eq:dustmass}.

Finally, we compute the $q$ parameter for the far--IR/radio correlation as defined by \citet{helou1985}:
\begin{equation}
\label{eq:fir.radio}
q = {\rm log}\left (\frac{L_{\rm 40-120\mu m}}{\rm 3.75\times 10^{12}\, W}\right )-{\rm log} \left(\frac{L_{\rm 1.4 GHz}}{\rm W\,\, Hz^{-1}}\right)
\end{equation}
where $L_{\rm 40-120\mu m}$ is the integrated IR luminosity between 40 and 120\micronend, and $L_{\rm 1.4GHz}$ is the rest--frame 1.4 GHz continuum radio luminosity.  Since our sources are at $z\sim 2$, their observed 1.4 GHz flux corresponds to a rest--frame frequency of 4.2 GHz.  Therefore, we applied a $k$--correction similar to that used in \citet{yang2007a}, assuming a power--law synchrotron $F_{\rm \nu,radio} \propto \nu^{-\alpha_r}$, where we assume $\alpha_r = 0.8$ as in typical local star--forming systems \citep{condon1983,condon1992}.  The results, as with the far--IR luminosity and dust properties, are listed in Table~\ref{tab:fits}, and discussed in the following sections.

\begin{table*}
\begin{tabular}{cccccccc}
\hline
\hline
Name & $z$ & $T_d$ & $L_{\rm 40-1000\mu m}$ & $L_{\rm 40-120\mu m}$ & $M_d$ &  $q$ \\
& & [K] & [$10^{12} L_\odot$] & [$10^{12} L_\odot$] & [$10^{8} M_\odot$] & \\
\hline
egs1 & 1.88 & $36\pm 6$ & $2.1\pm 0.9$ & $1.7\pm 0.7$ & $3.6\pm 1.6$ & $2.15\pm 0.20$ \\
egs4 & 1.90 & $37\pm 3$ & $3.3\pm 0.7$ & $2.7\pm 0.6$ & $4.4\pm 0.9$ & $2.29\pm 0.11$ \\
egs10 & 1.94 & $47\pm 3$ & $5.5\pm 0.7$ & $5.0\pm 0.7$ & $2.1\pm 0.3$ & $2.44\pm 0.08$ \\
egs11 & 1.80 & $41\pm 6$ & $2.5\pm 0.8$ & $2.2\pm 0.7$ & $1.8\pm 0.6$ & $2.25\pm 0.15$ \\
egs12 & 2.01 & $39\pm 3$ & $3.6\pm 0.8$ & $3.1\pm 0.8$ & $3.7\pm 0.8$ & $2.57\pm 0.15$ \\
egs14 & 1.86 & $41\pm 1$ & $9.7\pm 0.7$ & $8.3\pm 0.6$ & $8.3\pm0.6$ & $2.21\pm 0.04$ \\
egs21 & 3.01 & $62\pm 5$ & $8.9\pm 1.5$ & $8.4\pm 1.4$ & $0.8\pm 0.1$ & $2.32\pm 0.09$ \\
egs23 & 1.78 & $44\pm 3$ & $6.4\pm 0.6$ & $5.6\pm 0.6$ & $3.4\pm 0.3$ & $2.43\pm 0.06$ \\
egs24 & 1.85 & $43\pm 14$ & $1.2\pm 0.8$ & $1.1\pm 0.7$ & $0.7\pm 0.5$ & $2.07\pm 0.30$ \\
egs26 & 1.80 & $39\pm 4$ & $2.6\pm 0.6$ & $2.2\pm 0.5$ & $2.8\pm 0.6$ & $2.10\pm 0.11$ \\
egs24a & 2.12 & $34\pm 4$ & $4.0\pm 1.5$ & $3.1\pm 1.1$ & $9.5\pm 3.3$ & $2.03\pm 0.15$ \\
egsb2 & 1.59 & $34\pm 3$ & $2.3\pm 0.6$ & $1.8\pm 0.4$ & $5.6\pm 1.3$ & $2.00\pm 0.11$ \\
\hline
\end{tabular}
\caption{Derived parameters for our high redshift ULIRG sample, including: the mid--IR spectroscopic redshift (from Huang et al. in preparation) and the effective dust temperature ($T_d$), far--IR luminosity from 40--1000\micron ($L_{\rm 40-1000\mu m}$) and 40--120\micron ($L_{\rm 40-120\mu m}$), dust mass ($M_d$), and far--IR/radio $q$ parameter \citep[see \S~\ref{sec:fir.radio} and][]{condon1992}.  For details on the fitting procedure, see \S~\ref{sec:fit} and \citet{blain2003}.}
\label{tab:fits}
\end{table*}

\section{Dust Properties}
\label{sec:dust.properties}

Our far--IR and millimetre photometry that constrains both sides of the cold dust peak allows a determination of the far--IR luminosity, dust temperature, and dust mass from the far--IR alone.  As a result, the dust properties in our high redshift ULIRG sample can be determined with a minimum of additional assumptions.  While the properties of dust in these extreme environments and at high redshift is of itself an interesting result, it is particularly so in the context of analogous and related objects at different redshifts and with different selection techniques.  Therefore, we assembled results from five samples with extensive far--IR observations from the literature for comparison: another {\rm Spitzer}--selected high redshift sample (HZS), a low redshift sample (LZS), intermediate redshift sample (IZS), a submillimetre selected sample (SMS), and finally a sample of quasars -- which are thought to follow the ULIRG phase in an evolutionary sequence \citep{sanders1988a,sanders1988b,sanders1988c,hopkins2006,hopkins2007a,hopkins2007b} -- with extensive far--IR observations (QS).  Our far--IR SED parameterization is consistent with those used for the analysis of all samples discussed below -- including the HZS, IZS, SMS, and QS -- and therefore there are no hidden systematic biases.  

The HZS consists of bright, obscured 24\micron sources sected on the basis of their $R(24,0.6)$ ratio \citep{houck2005,yan2005,weedman2006a,weedman2006c}, which have mid--IR spectrscopic \citep{yan2007,sajina2007} and MAMBO 1200\micron photometric followup \citep{lutz2005,sajina2008}.  As we note in \S~\ref{sec:results}, \citet{lutz2005} find a signficantly lower average $\tilde{R}(1200,24)$ ratio than the objects in our sample, even among the most obscured ($R(24,0.6)\geq 1.5$) subsample.  The authors find a median dust temperature of $32\pm 8$ K ($35\pm 8$ K for 14 strong--PAH srouces, and $32\pm 8$ K for 34 weak--PAH sources), as compared to $41\pm 5$ K for our sample of 12 objects, with comparable far--IR luminosities and dust masses.  This difference may arise from the nature of their sources, combined with their SED fitting technique; the HZS has shown a significant contribution to the far--IR by an AGN \citep[see also \S~\ref{sec:fir.radio}][]{weedman2006c,sajina2008}, which the authors account for by fitting a combination of models to the mid--IR \citep[see][]{sajina2007} and far--IR separately in order to separate out the cold dust population.  However, it is most likely driven largely by their choice of $\beta=2$, which leads to cooler fitted dust temperatures than the $\beta=1.5$ adopted here \citep{blain2003,sajina2006} -- indeed, when we adopt $\beta=2.0$ we find a median dust temperature of $33\pm 4$ K while the total IR luminosity decreases by $\lsim 15 \%$.  Furthermore, since their mid--IR models will implicitly include some of the warm dust heated by the starburst, their two--component fitting method may also lead to somewhat lower dust temperatures.  Therefore, while theirs is a rather different population of objects, we find that the dust properties in both are similar.  This is an interesting result, considering the large fraction of obscured AGN in such samples \citep{weedman2006c,sajina2008}; the cold dust appears to be similar in both AGN and starburst dominated systems.

The LZS is the SCUBA Local Universe Galaxy Survey \citep[SLUGS][]{dunne2000}, which consists of 104 local ($z\lsim 0.05$) infrared luminous galaxies selected from the 60\micron flux limited {\em IRAS} Bright Source Catalog \citep[BSC:][]{soifer1989} with complete 450 and 850\micron followup with the Submillimetre Common User Bolometer Array \citep[SCUBA:][]{holland1999} at the James Clerk Maxwell Telescope (JCMT).  The IZS \citep{yang2007a} was drawn from the FSC--FIRST Survey \citep{stanford2000}, which consists of a matched sample of 60\micron {\em IRAS} Faint Source Catalog \citep[FSC:][]{moshir1992} and 1.4 GHz sources from the Faint Images of the Radio Sky at Twenty Centimeters \citep[FIRST:][]{becker1995} Survey, with complete 350\micron followup from the SHARC--II camera \citep{dowell2003} at the Caltech Submillimetre Observatory (CSO).  The SMS \citep{kovacs2006} consists of SMGs identified at 850\micron in blank field SCUBA surveys with 20cm radio counterparts, optical redshifts \citep{chapman2005}, and complete 350\micron followup with SHARCH-II \citep[for a similar sample, see][]{coppin2008.sharcii}.  Finally, the QS \citep{haas2000,haas2003} consists of 64 quasars selected from the Palomar Green Survey \citep[PG:][]{schmidt1983} with far--IR coverage from 5--200\micron from ISOPHOT \citep{lemke1996} on board the {\em ISO} satellite and some (sub)millimetre coverage with MAMBO and SCUBA.  

\begin{figure}
\epsfig{figure=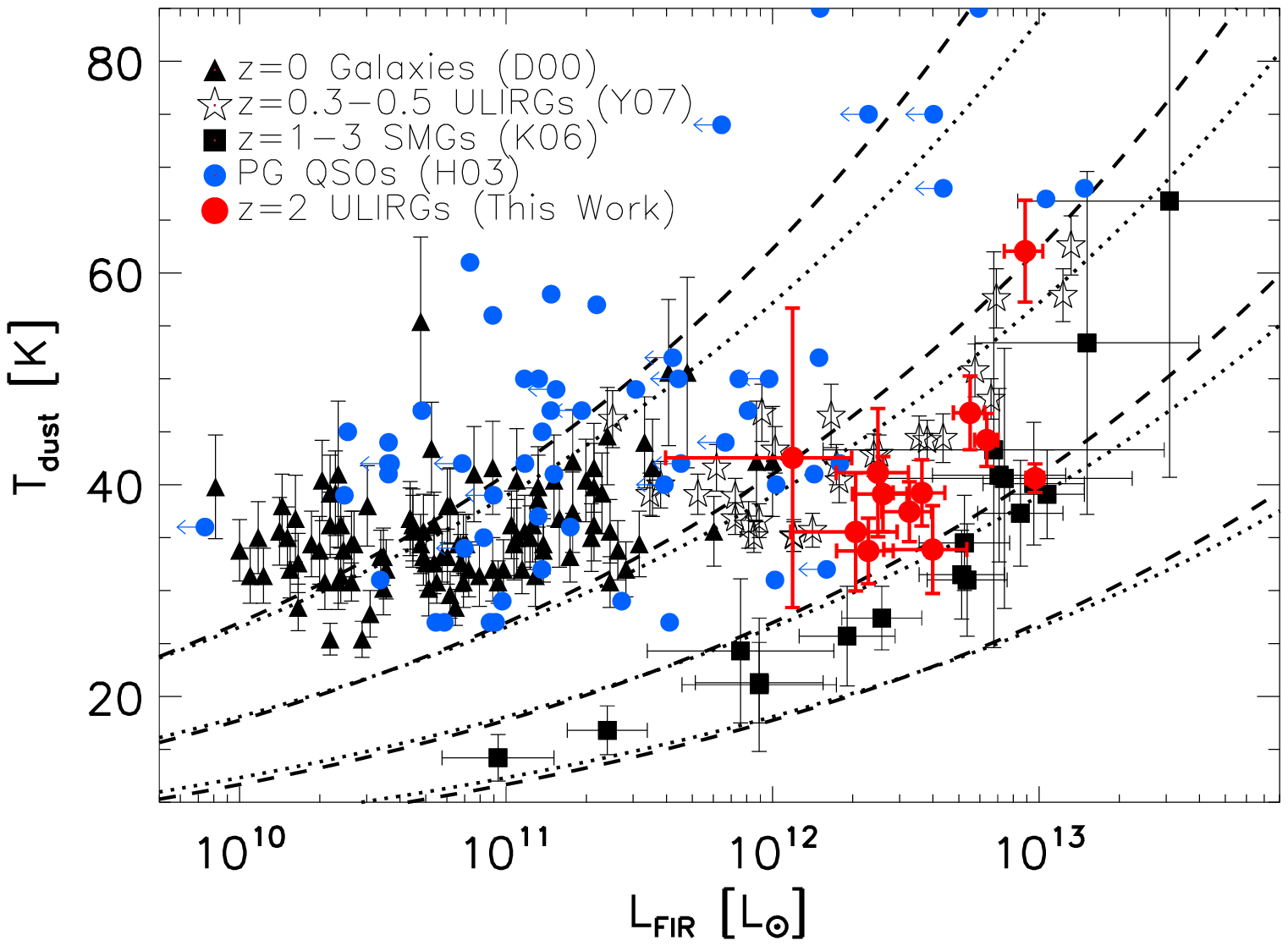,width=79mm}
\caption{The temperature--luminosity relation $L_{FIR}-T_d$, where $L_{FIR}$ is taken from 40-1000\micron.  The fitting method is detailed in \S~\ref{sec:fit}, and the values for our sample tabulated in Table~\ref{tab:fits}.  Included are results from our sample (red circles) as compared to local \citep[D00:][black triangles]{dunne2000}, intermediate redshift \citep[Y07:][open stars]{yang2007a}, high redshift submillimeter galaxies \citep[K06:][black squares]{kovacs2006}, and PG quasars \citep[H03][blue circles]{haas2003}.  The dashed lines indicated tracks in constant dust mass (see Equation~\ref{eq:lfir.md} and \S~\ref{sec:dust.properties}) with (left to right) $M_d = 10^{7}$, $10^{8}$, $10^{9}$, and $10^{10}$ $M_\odot$, assuming a fixed emissivity of $\beta=1.5$.  The dotted lines are the same for $\beta=2.0$.}
\label{fig:temp.vs.lum}
\end{figure} 

\begin{figure}
\epsfig{figure=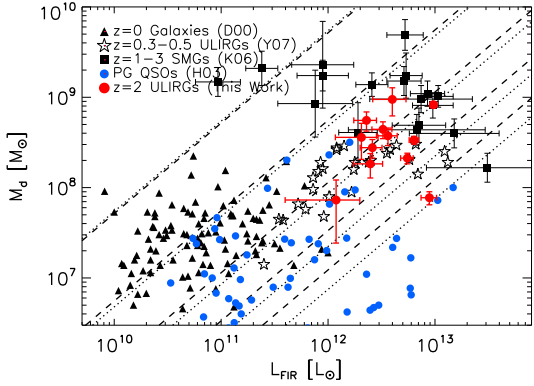,width=79mm}
\caption{Same Figure~\ref{fig:temp.vs.lum} for the luminosity--dust mass $L_{FIR}-M_d$ relation.  The dashed lines are tracks in constant dust temperature with (left to right) $T_d = 20$, 30, 40, 50, and 60 K.}
\label{fig:md.vs.lum}
\end{figure} 

\begin{figure}
\epsfig{figure=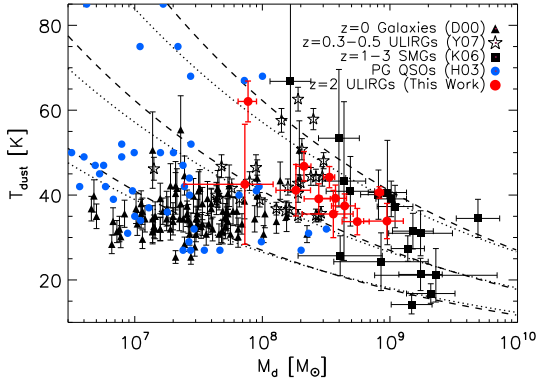,width=79mm}
\caption{Same Figure~\ref{fig:temp.vs.lum} for the luminosity--dust mass $L_{FIR}-M_d$ relation.  The dashed lines are tracks in constant far--IR luminosity with (left to right) $L_{FIR} = 10^{11}$, $10^{12}$, and $10^{13}$ $L_\odot$.}
\label{fig:md.vs.td}
\end{figure} 

For single temperature dust models, the total far--IR luminosity $L_{FIR}$ can be related to the dust mass $M_d$ and temperature $T_d$ via the dust mass absorption coefficient $\kappa_\lambda$ \citep{debreuck2003,yang2007a}: 
\begin{equation}
L_{FIR} = 4\pi M_d \int \kappa_\lambda B_\nu(\lambda, T_d) d\nu
\end{equation}
where $B_\nu$ is the thermal dust spectrum at rest--frame wavelength $\lambda$.  For greybody thermal emission, this yields:
\begin{equation}
\label{eq:lfir.md}
L_{FIR} = \frac{8 \nu h}{c^2} \frac{\lambda^\beta \kappa_\lambda}{c^\beta} \left ( \frac{k T_d}{h} \right)^{4+\beta} \Gamma(4+\beta) \zeta (4+\beta) M_d
\end{equation}
where $\beta$ is the dust emissivity, $\Gamma$ is the gamma function, and $\zeta$ is the Riemann Zeta function.   As a toy model, we use the Milky Way dust model of \citet{weingartner2001} at a reference wavelength of 125\micron ($\kappa_{\rm 125\mu m} \approx 17$ g$^{-1}$ cm$^{2}$), and make the assumption that $\beta = 1.5$ and does not vary significantly between objects -- which is approximately the case locally \citep[e.g.,][]{dunne2000,klaas2001,yun2002,yang2007b}.  Also, for all the following comparisons the total far--IR luminosity is taken from 40--1000\micron ($L_{FIR} = L_{\rm 40-1000\mu m}$) unless otherwise stated.

In Figure~\ref{fig:temp.vs.lum} we show the $L_{FIR}-T_d$  relation for our sources as compared to the LZS, IZS, SMS, and QS.  The dashed lines represent tracks (left to right) in constant $M_d = 10^7$, $10^8$, $10^9$ and $10^{10}$ $M_\odot$.   We find the objects in our sample all fall within a relatively narrow range of $35\lsim T_d\lsim 45$, though they vary by an order of magnitude in $L_{FIR}$.  This is similar to both the LZS and IZS, but characteristically different from the SMS, which resembles a constant $M_d$ selection with a broad $T_d$ distribution extending to lower temperatures.  The QS, by contrast, does not follow a clear trend in $L_{FIR}$, thought those with high luminosities $\gsim 10^{11} L\odot$ tend to have higher dust temperatures and lower dust masses than U/HyLIRGs of comparable luminosity.

This is further illustrated in Figures~\ref{fig:md.vs.lum} and \ref{fig:md.vs.td}, where we present the $M_d-L_{FIR}$ and $T_d-M_d$ scalings respectively.  In Figure~\ref{fig:md.vs.lum}, the dashed lines represent tracks in constant (left to right) $T_d = 30$, 40, 50, and 60 K, while in Figure~\ref{fig:md.vs.td} they represent tracks (left to right) in constant $L_{FIR}  = 10^{11}$, $10^{12}$, and $10^{13}$ $L_\odot$.  These two projections also show the dust selection function of the different samples samples, whereby the LZS (roughtly), IZS and our sample follow a track in constant temperature, the SMS is constant in dust mass, and the QS is preferentially warmer and high luminosity with lower dust mass.  

\section{The Far--IR/Radio Correlation at High Redshift}
\label{sec:fir.radio}

It has long been recognized that there exists a tight correlation between the far--IR and synchrotron radio emission in star--forming galaxies, spanning nearly four decades in luminosity \citep{helou1985,condon1991,condon1992,yun2001}.  Though the origin of this correlation is an active subject of ongoing research, the popular wisdom states that it arises from a ``cosmic conspiracy" that fixes the ratio of synchrotron losses due to magnetic fields associated with massive star formation and inverse--Compton from the ambient radiation field \citep{hummel1986,hummel1988,condon1992}.  This hypothesis is apparently born out in observations of local starbursts, which show a roughly constant far--IR/radio correlation over four decades in magnetic field energy density \citep{condon1991c}.  

However, should the far--IR/radio correlation arise from this fixed ratio of synchrotron and inverse--Compton losses, one might expect it to evolve with redshift.  The physics that govern this ratio are complex and not entirely understood \citep[see e.g.,][]{voelk1989,thompson2006}.  Therefore, it is important to check if it remains valid and/or constant at high redshift and in more luminous systems.

At intermediate redshift $0.3\lsim z \lsim 1$, there is some evidence for the constacy of the far--IR/radio correlation \citep{garrett2002,gruppioni2003,appleton2004,boyle2007}.  However, at high redshift $z\gsim 1.5$, the picture is somewhat less clear \citep{kovacs2006,vlahakis2007,ibar2008,sajina2008}, and the effects of sample selection -- in particular, contamination by radio AGN -- are uncertain.  Our sample offers an attractive opportunity to probe the far--IR/radio correlation in a relatively narrow redshift range, and for a well--defined sample.  

\begin{figure}
\epsfig{figure=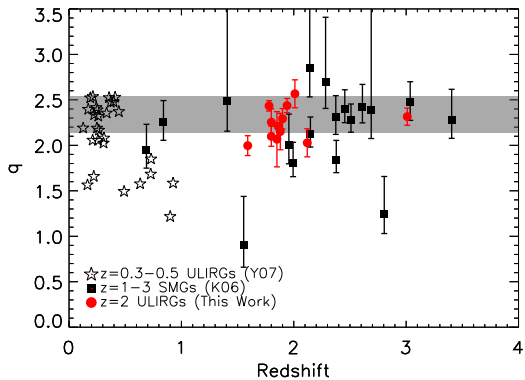,width=79mm}
\caption{The evolution of the far--IR/radio correlation with redshift parameterized by $q$ (see Equation~\ref{eq:fir.radio} and \S~\ref{sec:fit} for definitions).  The labeling is the same as Figure~\ref{fig:temp.vs.lum}, and the grey shaded region indicates the local value and associated dispersion from \citet{yun2001}.}
\label{fig:q.vs.z}
\end{figure} 

\begin{figure}
\epsfig{figure=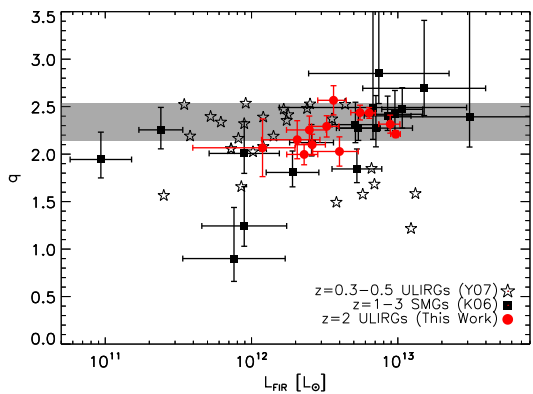,width=79mm}
\caption{The correlation between the far--IR/radio correlation coefficient $q$ (see Equation~\ref{eq:fir.radio} and \S~\ref{sec:fit} for definitions) with the far--IR luminosity $L_{FIR} = L_{\rm 40-1000\mu m}$.  The labeling is the same as Figure~\ref{fig:temp.vs.lum}, and the grey shaded region indicates the local value and associated dispersion from \citet{yun2001}.}
\label{fig:q.vs.lum}
\end{figure} 

Overall, our sample has a mean $<q> = 2.23\pm 0.04$ (see Equation~\ref{eq:fir.radio}) with intrinsic dispersion $\sigma_q = 0.19$ -- well in line with results for local systems.  In fact, only one object (egsb2) is inconsistent with the local $q$, and it is known to harbor an X--ray luminous AGN (Huang et al. in prepration).  This supports the conclusion of Huang et al. (2008, in preparation) that AGNs contribute little to the bolometric luminosity of these objects.  Furthermore, it is substantially different from the resutls of \citet{sajina2008} for obscured 24\micron sources, which cover a wide range in $q$ both for strong and weak PAH emitters -- the mean $<q> = 1.90\pm 0.13$ with $\sigma_q =  0.34$ for their ``strong--PAH" sources with radio detections, and $<q> = 1.38\pm 0.10$ with $\sigma_q = 0.47$ for their ``weak--PAH" sources.  This is, however, not surprising considering this sample selection is known to be biased towards obscured AGN \citep{weedman2006c}.  We take this as evidence to suggest that our selection criteria much more efficiently picks starburst--dominated systems in a narrow redshift range.  

The radio/far--IR properties of this sample are also consistent with the AGN contribution inferred from their X--ray/far--IR properties.  The rest--frame 25--60\micron flux density ratio versus $L_X/L_{IR}$ -- where $L_X$ is the rest--frame full band 2--10 keV X--ray luminosity -- has been shown to be sensitive to AGN activity \citep{deGrijp1985,deGrijp1987,risaliti2000}.  Huang et al. (in preparation) find that only three objects in this sample have even marginal $2\sigma$ X--ray detections: egs1 ($L_X \approx 10^{42}$ \ergspers, $L_X/L_{IR}\approx 10^{-3}$, $L_{\rm 25\mu m}/L_{\rm 60\mu m}\approx 0.2$), egs14 ($L_X \approx 10^{42}$ \ergspers, $L_X/L_{IR}\approx 3\times 10^{-4}$, $L_{\rm 25\mu m}/L_{\rm 60\mu m}\approx 0.06$), and egsb2 ($L_X \approx 10^{43}$ \ergspers, $L_X/L_{IR}\approx 10^{-2}$, $L_{\rm 25\mu m}/L_{\rm 60\mu m}\approx 0.15$).  Of these, only egsb2 is consistent with a significant contribution from an obscured AGN \citep[see Figure 5 of][]{risaliti2000}.  Also, while egs24 is not detected in the X--rays, its 25--60\micron flux density ratio suggests the presence of a Compton--thick AGN \citep[$N_H > 4\times 10^{24}$ cm$^{-2}$;][]{risaliti2000}.  The remaining 8 objects in our sample have a stacked X--ray flux corresponding to $<L_X> \sim 10^{42}$ \ergspers, along with $<L_{IR}> = 5\times 10^{12}\, L_\odot$ ($<L_X>/<L_{IR}> \lsim 5\times 10^{-4}$) and $<L_{\rm 25\mu m}/L_{\rm 60\mu m}> \approx 0.15$, which are all consistent with a pure starburst.

In Figures~\ref{fig:q.vs.z} and \ref{fig:q.vs.lum} we present the $q$ scaling with redshift and far--IR luminosity $L_{FIR}$  for our sample (red circles), local systems \citep[grey shaded region with $q = 2.34\pm 0.2$;][]{yun2001}, the IZS (open stars), and the SMS (black squares).  Immediately it appears that the local far--IR/radio correlation applies over a wide redshift range $0.2\lsim z\lsim 3$ over more than two decades in luminosity.  The apparent inverse trend of $q$ with both $L_{FIR}$ and redshift in the IZS is probably due to their sample selection, which is more likely to pick out IR bright AGN at higher redshifts (and luminosities).   By contrast, both our sample and the SMS show a clear trend towards higher $q$ at higher $L_{FIR}$: the Spearman rank correlation coefficients are $\rho = 0.52$ at $\sim 2\sigma$ and $\rho = 0.79$ at $\sim 3\sigma$ for our sample and the SMS respectively.  This trend is not seen in surveys of local systems spanning $10^{10}L_\odot \lsim L_{FIR} \lsim 10^{12} L_\odot$ \citep{yun2001}.  However, it is not clear whether or not this effect is due to sample selection -- our sample and the SMS are selected in very different ways, and show different trends in dust properties (see \S~\ref{sec:dust.properties}) -- or the strong redshift evolution of the bright end of the galaxy IR luminosity function \citep{sanders1996,lefloch2005,hopkins2007a} and that of massive galaxies \citep[e.g.,][]{conselice2007b} versus the comparatively weaker evolution of flat--spectrum radio sources \citep{dunlop1990}.  Such a selection effect would potentially bias $z\gsim 1.5$ systems with luminosities of $L_{FIR}\sim 10^{10-11}$ to radio--loud AGN as star formation moves preferentially to higher luminosity objects.

\section{Conclusion}
\label{sec:conclusion}

We present far--IR observations, including 1200\micron photometry with MAMBO, of high redshift ULIRGs selected by Huang et al. (in preparation) on the basis of their observed mid--IR colour.  All the objects in the sample have mid--IR spectroscopic redshift and lie in a remarkably tight redshift window centered at $z\sim 2$.  We successfully detected (S/N $>3$) 9 of 12 total sources.  When combined with existing MIPS photometry at 70 and 160\micronend, this provides constraints on both sides of the cold dust peak, allowing a more robust determination of the far--IR SED.  

We fit a model to the photometry, from which we estimate both the far--IR luminosity and dust properties -- including temperature and mass.  When compare to other samples, including IR luminous systems at low and intermediate redshift, submillimetre selected galaxies, and QSOs, and find that our sample, along with the low and intermediate redshift sources, is roughly a constant dust temperature selection.  Submillimetre sources, by contrast, are more similar to a constant dust mass selection.  Finally, we use existing radio observations at 20cm to test the far--IR/radio correlation at high redshift.  We find that all but one of the sources in our sample is consistent with the local relation, suggesting that it remains valid at high redshift.  This furthermore indicates that the Huang et al. (in preparation) method efficiently selects massive, high redshift, starburst--dominated systems in a remarkably tight redshift range.

\section*{Acknowledgements}
Thanks to Lars Hernquist, Philip Hopkins, T. J. Cox, Dusan Keres, Chris Hayward, and Stephanie Bush for helpful discussions, and to the referee, Stephen Serjeant, for his helpful suggestions for improving this manuscript.  This work is based in part on observations made with the {\em Spitzer Space Telescope}, which is operated by the Jet Propulsion Laboratory, California Institute of Technology under a contract with NASA. Support for this work was provided by NASA through an award issued by JPL/Caltech.  This work includes observations made with IRAM, which is supported by INSU/CNRS (France), MPG (Germany) and IGN (Spain).  Thanks to the staff of IRAM Granada for their support in taking the MAMBO observations presented here.  

\bibliographystyle{apj}
\bibliography{../../smg}

\end{document}